# 100% reliable algorithm for second-harmonic-generation frequency-resolved optical gating

**RANA JAFARI,\* TRAVIS JONES, AND RICK TREBINO**

*School of Physics, Georgia Institute of Technology, 837 State Street, Atlanta, GA 30332, USA*
*\*rjafari7@gatech.edu*

**Abstract:** We demonstrate a novel algorithmic approach for the second-harmonic-generation (SHG) frequency-resolved optical gating (FROG) ultrashort-pulse-measurement technique that always converges and, for complex pulses, is also much faster. It takes advantage of the Paley-Wiener Theorem to retrieve the precise pulse spectrum—half the desired information—directly from the measured trace. It also uses a multi-grid approach, permitting the algorithm to operate on smaller arrays for early iterations and on the complete array for only the final few iterations. We tested this approach on more than 25,000 randomly generated complex pulses with time-bandwidth products up to 100, yielding SHG FROG traces to which noise was added, and have achieved convergence to the correct pulse in all cases. Moreover, convergence occurs in less than half the time for extremely large traces corresponding to extremely complex pulses.

## 1. Introduction

Ultrashort-laser-pulse measurement has remained challenging since the origin of the field of ultrafast optics. Intensity autocorrelation yields only a blurred record of the amplitude of the pulse and no phase information [1]. Worse, for pulse trains with pulse-shape instability, an ultrashort "coherent artifact" contaminates the autocorrelation trace, often inducing researchers to claim erroneously short pulse lengths [2, 3]. Modern techniques provide the complete temporal intensity and phase but still have serious drawbacks. Some measure *only* the coherent artifact [3-8]. Some can measure only simple pulses. Others are too complex experimentally or have limited applications.

Arguably, the most practical, reliable, and well-developed method is frequency-resolved optical gating (FROG), whose variations generate various types of spectrograms of the pulse, depending on the nonlinear-optical process involved [1]. FROG routinely measures pulses from attoseconds to nanoseconds in length [9, 10], from the XUV to the IR in wavelength [11, 12], and from simple to extremely complex in shape [13, 14]. Also, FROG's two-dimensional traces significantly overdetermine the pulse, so that discrepancies between measured and retrieved traces indicate (in)stability of a pulse train [5, 7].

The simplest, most sensitive, and most prevalent FROG variation uses second-harmonic generation (SHG). Indeed, the pulse-shaping community routinely uses SHG FROG to measure potentially very complex pulses used in nonlinear spectroscopy and in carrying out coherent anti-Stokes Raman scattering [15, 16]. Additionally, it has been proven mathematically that all [17] pulses can be uniquely determined by SHG FROG up to well-known trivial ambiguities [18, 19]. SHG FROG's iterative pulse-retrieval algorithm is, however, the least reliable. While it generally converges quickly (~100ms) and reliably (~100%) for simple pulses, it tends to stagnate for more complex pulses. For pulses with time-bandwidth products (*TBP*s) of 20, for example, the convergence probability falls to ~70%, and, for *TBP*s of 100, it falls to ~40% [14]. And the convergence times are long: ~1 minute and ~1/2 hour, respectively, when using code written in MATLAB.

Various alternative SHG FROG algorithms have been proposed. Genetic algorithms are more reliable (although how much so is unknown), but much slower [20]. Neural nets retrieve pulses quickly, but so far only for very simple pulses [21]. Other algorithms address other issues [22, 23]. Ptychographic algorithms address problems like missing data, rather than speed or reliability [24]. For other FROG geometries [25], simulated-annealing algorithms have been used but are also extremely slow [26]. So far, only the algorithm for XFROG (a simple spectrogram) has achieved 100% convergence [14, 27], but XFROG's requirement of a known gate pulse severely limits its applicability. In view of the prevalence and importance of SHG FROG for measuring ever more complex pulses and its need for reliable convergence to determine pulse-train stability, the development of a reliable algorithm for it is critical.

We have now solved this long-standing problem. Our solution uses any SHG FROG algorithm (we use the well-known generalized projections algorithm), but within a novel approach, using the Paley-Wiener Theorem, which states that the Fourier-transform of a function with compact support is analytic. This means that the Fourier transform of the pulse spectrum is continuous, as are all of its derivatives. This allows us to *directly retrieve the pulse spectrum from the measured trace and hence to generate a vastly improved initial guess*. It also uses multiple frequency and time grids, in which the algorithm operates initially on smaller arrays—and hence much faster—and on the complete array for only the final few iterations [27, 28]. In effect, this allows us to obtain a vastly improved initial guess for the pulse spectral intensity and phase before resorting to the full SHG FROG trace (the slowest step of any algorithm), thus requiring only a few iterations on it. We call this approach the Retrieved-Amplitude N-grid Algorithmic (RANA) approach.

## 2. FROG-trace marginals and the direct retrieval of the pulse spectrum

SHG FROG can be represented mathematically in the delay ($\tau$) and frequency ($\omega$) domains as:

$$I_{FROG}^{SHG}(\omega,\tau) = \left| \int_{-\infty}^{+\infty} E(t)\, E(t-\tau) \exp(i\omega t) dt \right|^2. \tag{1}$$

where $E(t) = \sqrt{I(t)} \exp[-i\phi(t)]$ is the complex amplitude of the pulse, and $I(t)$ and $\phi(t)$ are the intensity and phase as a function of time, $t$, respectively. In addition, the delay and frequency marginals, $M(\tau)$, and $M(\omega)$—the integrals of the trace over all frequencies or delays, respectively—have been used as self-consistency checks for measurements or correction techniques for traces with systematic error (see Fig.1).

$$M(\tau) \equiv \int I_{FROG}(\omega,\tau)\, d\omega \tag{2}$$

$$M(\omega) \equiv \int I_{FROG}(\omega,\tau)\, d\tau \tag{3}$$

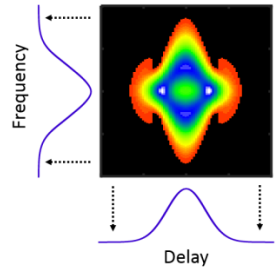

Fig.1. SHG FROG marginals.

The delay marginal is easily seen to be the pulse temporal intensity's autocorrelation, and a comparison of it and an independently measured autocorrelation provides confidence that the SHG FROG trace was correctly measured. Little else can be gleaned from the delay marginal, however, as retrieving the temporal intensity $I(t)$ from it, for example, is mathematically equivalent to the highly ill-posed one-dimensional phase-retrieval problem. Specifically, by the autocorrelation theorem, the Fourier-transform of the delay marginal is the squared magnitude of the Fourier-transform of the temporal intensity. In other words, all the phase information needed for this inverse Fourier transform is lost. Finding the correct phase information for this quantity is hopeless, as infinitely many ambiguous values—all possible values from 0 to $2\pi$ continuously—are possible at every delay point, and no method exists, or is likely to exist, for finding it.

The frequency marginal in SHG FROG, $M^{SHG}(\omega)$, which is analogously related to the pulse spectrum, $S(\omega)$, appeared to have similar ambiguity issues and so has also not been considered useful for pulse retrieval. Unlike the delay marginal, however, it corresponds to the *autoconvolution*, not the autocorrelation, of the spectrum of the pulse [1]. As a result, the (inverse) Fourier-transform of the frequency marginal is simply the square—not the *mag* square—of the inverse Fourier-transform of the spectrum $s(t)$:

$$\mathcal{F}^{-1}\left\{M^{SHG}(\omega)\right\} = \left(\mathcal{F}^{-1}\left\{S(\omega)\right\}\right)^2 \equiv s(t)^2. \tag{4}$$

This inversion problem is actually significantly simpler and much less ill-posed. The function $s(t)$ can then be trivially found in terms of the frequency marginal, $M^{SHG}(\omega)$:

$$s_\pm = \pm\sqrt{\mathcal{F}^{-1}\left\{M^{SHG}(\omega)\right\}} \tag{5}$$

where, of course, at every point in time, $t$, the square root has two complex roots, $s_\pm(t)$, where we define $s_+(t)$ to be the root with positive real component, and $s_-(t)$ to be the root with negative real component, corresponding to phase values of $\phi(t)$ and $\phi(t) + \pi$, respectively.

To better understand this inversion problem, suppose that the SHG FROG trace has $N$ delays and $N$ frequencies and so is an $N \times N$ array. The frequency marginal then has $N$ points. The above inversion then yields $2^N$ possible solutions, which is clearly infinitely less problematic than the problem of retrieving the intensity from the delay marginal with its infinite number of possible solutions at every point. But this still seems a difficult and ill-posed inversion problem.

Fortunately, the spectral-inversion problem enjoys several strong constraints. First is the fact that the spectrum is always square integrable, which follows from the facts that the pulse has finite energy and the spectrum is always positive. This implies, by the Paley-Wiener Theorem, that its inverse Fourier-transform $s(t)$ is a $C^\infty$ function, that is, is continuous and has infinitely many continuous derivatives at every time point [29]. See Fig. 2. Even better, additional information is provided by the obvious fact that the spectrum is real, so its inverse Fourier-transform's real and imaginary parts must be even and odd, respectively. Furthermore, the always-positive nature of the spectrum can also be used to eliminate any remaining ambiguities, although this only proves necessary in the presence of noise. In other words, retrieving the spectrum directly from the trace is always possible.

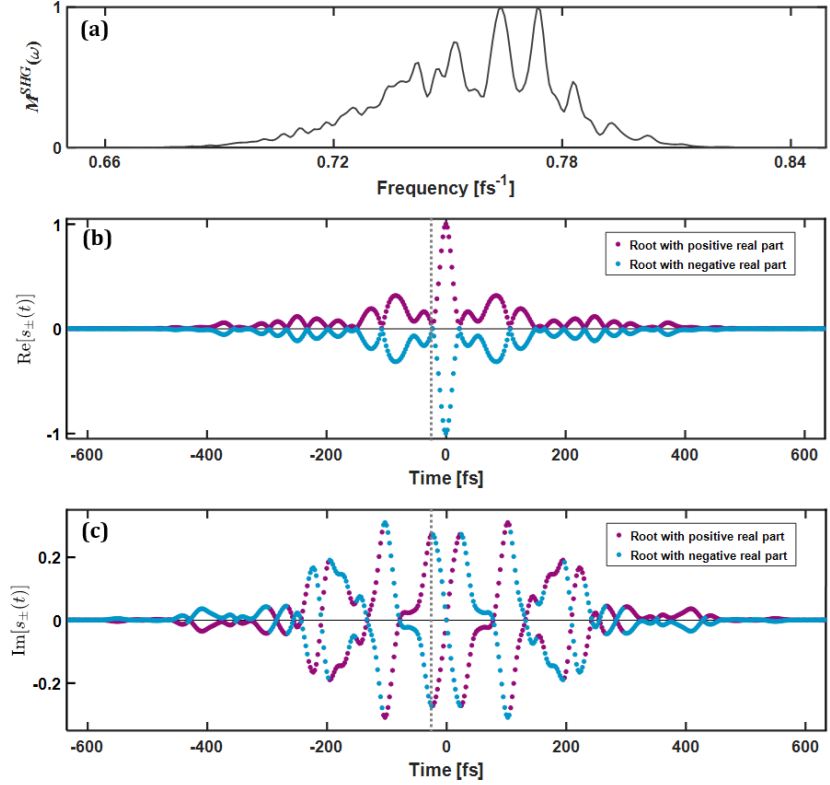

Fig. 2. (a) Frequency marginal of a SHG FROG trace, $M^{SHG}(\omega) = S(\omega)*S(\omega)$. Plots of (b) real, and (c) imaginary parts of $s_\pm(t) = \pm\sqrt{\mathcal{F}^{-1}[M^{SHG}(\omega)]}$ for a randomly generated complex pulse. The root with positive (negative) real component is shown in purple (blue), respectively. Note that the imaginary components of the roots can have either positive or negative values. As a result, the real part is always continuous, but the imaginary part may not be. But both components of the actual complex function $s(t)$ must be continuous. Therefore, the discontinuity of the imaginary part almost always determines sign of the correct roots when the real values are close to zero. Conversely, the continuity of real values can be applied when the imaginary points are close to zero. The gray dotted lines in the two plots pass in between temporal points $t = -25$fs and $t = -22.5$fs. If one begins at $t = 0$ and moves in the decreasing time direction on purple curves, when transitioning from $t = -22.5$fs to $t = -25$fs, the correct sign of the root must change to satisfy the $C^\infty$ condition. Analogous arguments apply to all of the derivatives of $s(t)$, and such higher order discontinuities are also apparent in the plots.

In practice, we find that simple continuity of $s(t)$ eliminates most, if not all, ambiguities. We simply begin at some point $t_i$, and choose one root, $s(t_i)$. Then we find the two possible roots for the next temporal point, $t_{i+1}$. If $|s(t_i)|$ and $|s(t_{i+1})|$ are both nonzero, one of the roots $s_\pm(t_{i+1})$ will necessarily be considerably closer to $s(t_i)$ than its additive inverse. So, we eliminate the more distant root, and it is then on to the next temporal point. Only when the real and imaginary components of the roots are both near zero does this approach fail, and, in the wings of the pulse, such failure is irrelevant.

When $s(t_i)$ and $s(t_{i+1})$ are both near zero, continuity of $s(t)$'s first derivative is implemented analogously. And if this fails, continuity of the second derivative can then be used. While this procedure can be continued ad infinitum, we have only gone as far as the second derivative, as the ambiguity in the sign was always resolved by stopping at this order for the set of sample pulses. Figure 3 provides a flowchart of this procedure.

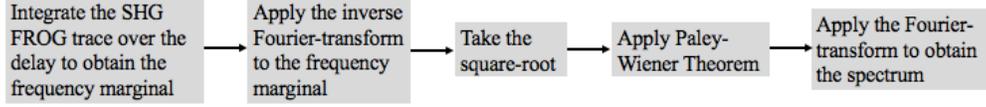

Fig.3. Schematic of steps for retrieving the spectrum of the pulse directly from the frequency marginal of its SHG FROG trace.

More quantitatively, we define the finite differences:

$$\Delta_{0\pm} = s_\pm(t_{i+1}) - s(t_i), \tag{6}$$

$$\Delta_{1\pm} = [s_\pm(t_{i+1}) - s(t_i)] - [s(t_i) - s(t_{i-1})], \tag{7}$$

and,

$$\Delta_{2\pm} = \{[s_\pm(t_{i+1}) - s(t_i)] - [s(t_i) - s(t_{i-1})]\} - \{[s(t_i) - s(t_{i-1})] - [s(t_{i-1}) - s(t_{i-2})]\}. \tag{8}$$

For $s(t)$ to be $C^\infty$, these quantities should all be as close to zero as possible. One could implement them sequentially, as described above, but we have instead defined weighted sums of the mag-squared differences $\Delta_{0\pm}$, $\Delta_{1\pm}$, and $\Delta_{2\pm}$ for both signs of root. We then compare their values at $t_{i+1}$ and choose the sign of $s_\pm(t_{i+1})$ that yields the lesser value of the mag-squared sums:

$$\varepsilon_\pm \equiv \alpha |\Delta_{0\pm}|^2 + \beta |\Delta_{1\pm}|^2 + \gamma |\Delta_{2\pm}|^2, \tag{9}$$

where we determined the weighting coefficients $\alpha$, $\beta$, and $\gamma$ empirically by optimizing the accuracy of the spectral amplitude retrieval on a sample set of simulated SHG FROG traces with scaled peak intensity of 1 using the rms difference between the retrieved and simulated spectrum with peak intensity of 1:

$$\Delta S = \sqrt{\frac{1}{N} \sum_{i=1}^{N} \left( S_i^{simulated} - S_i^{retrieved} \right)^2}. \tag{10}$$

The resulting values of these coefficients are 0.09, 0.425, and 1, respectively.

In implementing the above approach, we zero-padded the frequency marginal from length $N$ to $2N$ in order to improve the temporal resolution of $s(t)$. In addition, we only used half of the temporal range and implemented the required symmetry conditions for $s(t)$ to find the other half of $s(t)$. We used the other half of the temporal range for an additional, perhaps slightly different (in the presence of noise), estimate of $s(t)$. Finally, in the presence of noise, it is possible that this process will exhibit slight discrepancies, yielding one or more small negative regions in the spectrum $S(\omega)$. In this case, the magnitude of the resulting $S(\omega)$ can be used. Alternatively, one could re-evaluate some of the decisions made by the above algorithm and choose alternative solutions for $s_\pm(t_{i+1})$ for cases with the most similar values of $\varepsilon_\pm(t_{i+1})$ and choose instead the solution with the most positive-valued spectrum. In view of the fact that two slightly different spectra were retrieved (as discussed above), when negative values appeared, we simply chose the spectrum with fewer negative values (rejecting the other) and took the absolute value of that spectrum, but we mention the latter approach here for future possible consideration.

We find that retrieving the correct spectrum directly from the SHG FROG trace is easy and reliable: the only spectral ambiguities that occur are the trivial spectral ambiguities already known to be present in SHG FROG itself (e.g., spectra due to well-separated pulses). In the absence of noise, the above process always succeeded perfectly.

Of course, in practice, SHG FROG traces are usually contaminated with multiplicative noise (noise proportional to the intensity at that point). In the presence of such noise, the RANA approach yielded the correct spectrum more than 99% of the time. And when it failed, it still yielded a spectrum very close to the correct one, thus yielding an excellent initial guess for the iteration in all cases. As a result, this means that half of the pulse information can be found immediately, before the iterative phase-retrieval algorithm even begins its task, and *only the spectral phase remains to be found.*

So, we also find an excellent initial guess for the spectral phase. We do this by running the SHG FROG algorithm for reduced-size grids, using the directly retrieved spectrum. This quickly yields an excellent guess for the spectral phase before we resort to using the entire SHG FROG trace, which is the slowest step of any FROG algorithm.

This approach also works for other FROG variations, such as polarization-gate and transient-grating FROG, commonly used for measuring and shaping UV pulses and/or few-cycle pulses [30-32], which we describe in a different publication. The method for directly retrieving the spectrum is slightly different, however, because the expression for the frequency marginal is different [33].

## 3. Multi-grid iterative algorithm

To find the above mentioned initial guess for the spectral phase and to perfect the retrieved pulse, we take a cue from genetic algorithms, which begin with multiple initial guesses and systematically modify and cull them. In particular, all initial guesses that we use, use the above-retrieved spectrum, but with random spectral phases, and we simultaneously improve them using the standard generalized projections (GP) algorithm [1]. We also cull them, keeping only the best solutions as we proceed.

In addition, the use of multiple guesses naturally takes advantage of parallel processing, which has already found its way into phase-retrieval algorithms [34]. MATLAB provides an easy implementation of explicit parallelism, in which a single MATLAB command or M-function runs simultaneously on the available multiple cores. In this work, we used a four-core processor and MATLAB's parallel processing by simply executing a "par-for" (parallel for loop), and by doing so, four initial guesses are tried simultaneously.

Finally, our approach further takes advantage of multiple grids, in which we interpolate the trace to smaller sizes ($N/4 \times N/4$ and $N/2 \times N/2$) for the early iterations. If $N \times N$ is the size of the SHG FROG-trace array, the number of operations in an iteration scales as $N^2$. Thus, we test multiple guesses much more quickly on these smaller arrays and only retain the pulses whose traces best match the measured one. Also, we re-apply the retrieved spectrum from the frequency marginal if it provides a trace with a better delay marginal than the one obtained from the trace of the retained field. By comparing these autocorrelations, not only can we check whether the guess's current spectrum is a better one to be used, but we also ensure that the spectral phase matches the spectrum for a given value of $\omega$, as the first-order temporal phase ambiguity yields a shift (half of the spectral range) in the spectral domain. By the time the algorithm reaches the full array, only the few best pulses remain. There, they are tried one by one, and, nearly always, the first one converges to the correct trace in only a few iterations.

This is the motivation behind the name we chose for this approach: Retrieved-Amplitude N-grid Algorithmic (RANA) approach. The RANA approach implemented here uses the parameters presented in Table 1 for the number of initial guesses (IGs) and iterations on each grid for seven sets of pulses with different complexities.

Table 1. Parameters of the RANA approach used here. The value of the *G* error (rms difference between the measured and retrieved traces) for determining convergence, number of initial guesses (IGs), and number of iterations that are used on each grid in RANA approach for pulses with *TBP*s of 2.5 to 100.

| Pulse TBP | Array size, N | N/4 × N/4 | | N/2 × N/2 | | N × N | | Maximum G error | # of sample pulses |
|---|---|---|---|---|---|---|---|---|---|
| | | # of IGs | # of iterations | # of IGs | # of iterations | # of IGs | # of iterations | | |
| 2.5 | 64 | 12 | 25 | 8 | 20 | 4 | NA | 9.0e-3 | 5000 |
| 5 | 128 | 16 | 25 | 12 | 20 | 4 | NA | 6.4e-3 | 5000 |
| 10 | 256 | 32 | 25 | 16 | 25 | 4 | NA | 4.5e-3 | 5000 |
| 20 | 512 | 36 | 25 | 16 | 25 | 4 | NA | 3.2e-3 | 5000 |
| 40 | 1024 | 44 | 30 | 20 | 25 | 8 | NA | 2.2e-3 | 5000 |
| 80 | 2048 | 48 | 30 | 24 | 25 | 8 | NA | 1.6e-3 | 500 |
| 100 | 4096 | 60 | 30 | 32 | 25 | 8 | NA | 1.1e-3 | 200 |

## 4. Results

We tested the RANA approach on simulated sets of pulses with rms *TBP*s of 2.5, 5, 10, 20, 40, 80, and 100. To generate the simulated fields, we began with a random complex array in the time domain and multiplied it by a Gaussian temporal intensity profile. Then, we multiplied it in the frequency domain by a Gaussian spectrum. To obtain the desired *TBP*, we adjusted the widths of the temporal and spectral Gaussian intensities. The SHG FROG trace sizes of these pulses were chosen such that the intensity at the perimeter of the trace falls to 0.0001 of the peak intensity of the trace. The rms *TBP* of a pulse can be found directly from the SHG trace, but here we use the following more traditional expression to determine the rms *TBP*:

$$TBP = \frac{\int_{-\infty}^{+\infty} t^2 I(t)\,dt - \left(\int_{-\infty}^{+\infty} t\, I(t)\,dt\right)^2}{\int_{-\infty}^{+\infty} I(t)\,dt} \times \frac{\int_{-\infty}^{+\infty} (\omega-\omega_0)^2 S(\omega-\omega_0)\,d\omega - \left(\int_{-\infty}^{+\infty} (\omega-\omega_0) S(\omega-\omega_0)\,d\omega\right)^2}{\int_{-\infty}^{+\infty} S(\omega-\omega_0)\,d\omega} \quad (11)$$

Because SHG FROG traces are usually contaminated with multiplicative noise, 0.5% multiplicative noise was applied to each pixel of the trace to simulate pixel-to-pixel gain variations:

$$I_{FROG}^{noisy}(\omega_i, \tau_j) = I_{FROG}(\omega_i, \tau_j)\left(1 + m_{ij}\alpha\right) \quad (12)$$

where $m_{ij}$ is a pseudorandom number drawn from a zero mean unit-variance normal distribution, $\alpha$ is the noise fraction and the FROG trace has the peak intensity of one [14]. We reiterate here that noise in the trace could, in principle, and can, in practice, interfere with the direct retrieval of the spectrum from the frequency marginal. The spectral intensity retrieved from these noisy traces was actually slightly visibly different from the precise spectrum for a small fraction (<1%) of the traces (see Table 2 and Fig. 4). But the resulting spectrum, even when "wrong", differed only slightly from the actual spectrum and still provided an excellent initial guess and yielded rapid convergence to the correct pulse.

**Table 2. Performance of spectrum retrieval from the frequency marginal of the FROG traces with 0.5% multiplicative noise.**

| Pulse TBP | Array size, N | Average ΔS | # of pulses with rms error > 0.12 | # of sample pulses |
|---|---|---|---|---|
| 2.5 | 64 | 0.0056 | 3 out of 5000 | 5000 |
| 5 | 128 | 0.0080 | 16 out of 5000 | 5000 |
| 10 | 256 | 0.0110 | 17 out of 5000 | 5000 |
| 20 | 512 | 0.0151 | 25 out of 5000 | 5000 |
| 40 | 1024 | 0.0180 | 43 out of 5000 | 5000 |
| 80 | 2048 | 0.0227 | 2 out of 500 | 500 |
| 100 | 4096 | 0.0226 | 1 out of 200 | 200 |

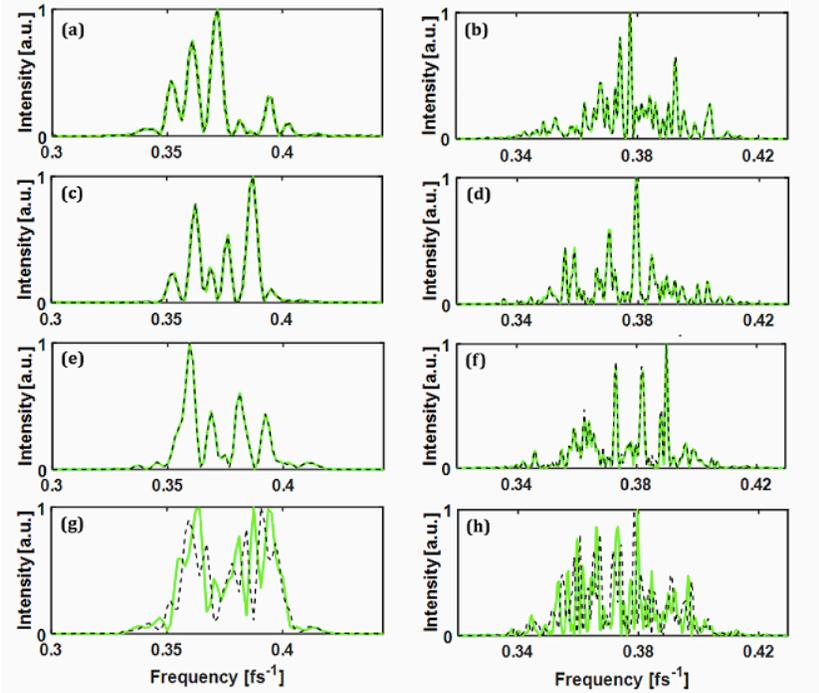

Fig. 4. Examples of spectra retrieved directly from noisy SHG FROG trace frequency marginals in our study. The simulated spectrum and the retrieved spectrum from the frequency marginal are shown in solid green and dashed black lines, respectively. First column: *TBP* = 5, and second column: *TBP* = 20. (a,b) The result corresponding to value ΔS in first quartile, (c,d) second quartile, and (e,f) third quartile, respectively. (g) ΔS = 0.17, and (h) ΔS = 0.16. Note, however, that these "incorrect spectra" still retain the rough structure of the spectrum and still provided excellent initial guesses for the iteration and yielded convergence to the correct pulse once the full array was used.

It should also be noted, as mentioned earlier, that the same spectral ambiguities that appear in SHG FROG measurement will also be present in the process of retrieving the spectral amplitude from $M^{SHG}(\omega)$. For instance, the ambiguity in the relative phase of sub-pulses in a double pulse results in different spectral intensities having the same FROG trace, and hence, the same spectrum autoconvolution [1]. This issue also appeared in a number of cases in our study, and hence the retrieval procedure generated the other possible spectrum. These cases cannot, of course, be considered algorithm stagnations, and we dealt with this issue by using a convergence condition involving the $G$ error (the rms difference between retrieved traces). We also confirmed that the retrieved pulses also agreed with the generated ones up to this ambiguity, but this was probably not necessary in view of the SHG FROG uniqueness proof.

Figure 5 shows a typical complex theoretical pulse with a *TBP* of 20 retrieved by the RANA approach.

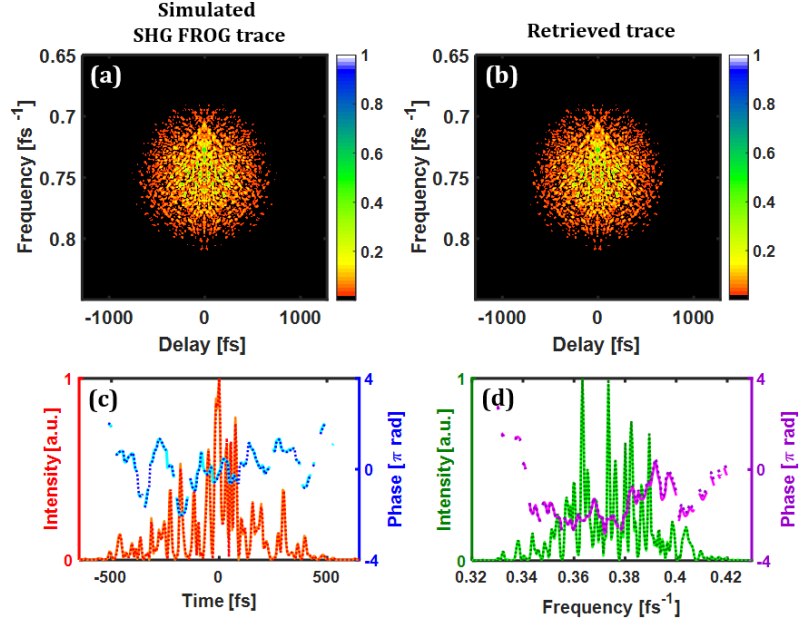

Fig. 5. Typical SHG FROG pulse-retrieval result with a $G$ error = 0.0018 and a trace size of 512 × 512 for a pulse with *TBP* = 20 and 0.5% multiplicative noise. (a) The simulated trace. (b) The retrieved trace. (c,d) The simulated temporal/spectral intensity and phase are shown in orange/light green and cyan/magenta, respectively. The retrieved temporal/spectral intensity and phase are shown in dashed red/dark green and dashed dark blue/dark purple, respectively.

Table 3 shows the results of running the RANA approach for thousands of pulses with *TBP*s as high as 100. For the statistical analysis of the convergence for pulses with *TBP* = 80 and 100, smaller samples of pulses were used for the sake of time. But, in view of the fact that convergence was nearly always achieved for the first pulse on the final $N \times N$ array even for these very complex pulses, the results are likely not impacted by this fact.

Table 3. Performance of GP algorithm and RANA approach on SHG traces with different complexity and size.

| Pulse TBP | Array size, $N$ | Average retrieval time for converging initial guess | GP algorithm percentage of convergence on first initial guess | # of sample pulses | Average retrieval time | RANA approach percentage of convergence on first initial guess |
|---|---|---|---|---|---|---|
| 2.5 | 64 | 0.051 s | 75% | 5000 | 0.150 s | (5000/5000) 100% |
| 5 | 128 | 0.30 s | 74% | 5000 | 0.289 s | (5000/5000) 100% |
| 10 | 256 | 1.29 s | 70% | 5000 | 1.25 s | (5000/5000) 100% |
| 20 | 512 | 6.03 s | 63% | 5000 | 4.49 s | (5000/5000) 100% |
| 40 | 1024 | 45.7 s | 56% | 5000 | 28.6 s | (5000/5000) 100% |
| 80 | 2048 | 315 s | 51% | 500 | 146 s | (500/500) 100% |
| 100 | 4096 | 31 min | 54% | 200 | 14 min | (200/200) 100% |

*The RANA approach never stagnated.*

## 5. Experimental results

To test the RANA approach with a relatively challenging *experimental* trace, we measured a SHG FROG trace of a chirped double-pulse using a Swamp Optics GRENOUILLE (model 8-50). See Fig. 6. We used the standard GP algorithm and also the RANA approach to retrieve the pulse for ten different sets of initial guesses. We found that the standard GP algorithm converged to the lowest $G$ error = 0.014 in only five of the ten attempts, while the RANA approach converged to this minimum error in all the tries.

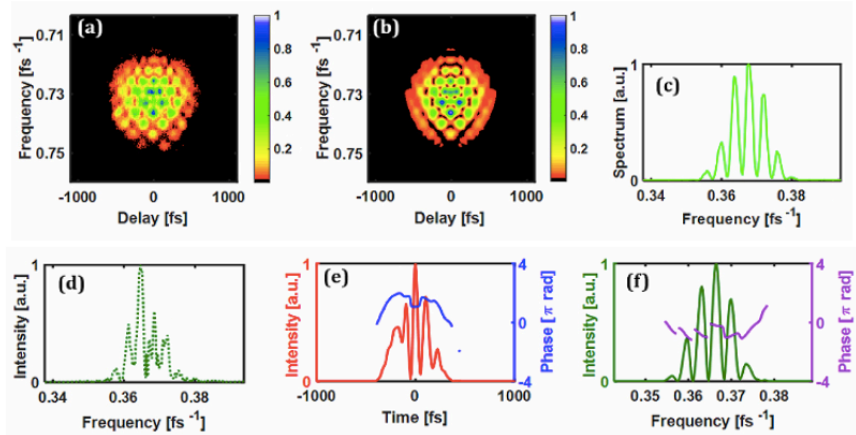

Fig. 6. (a) Experimental SHG FROG trace of a pulse with *TBP* = 4.1 used to study the performance of RANA approach. (b) The retrieved trace. (c) The spectrum measured by the spectrometer. (d) The spectrum retrieved directly from the SHG trace. Note that this spectrum actually differs slightly from the correct spectrum, but this does not affect the results. A 512 × 512 trace is used to obtain better temporal and spectral resolutions for this complex pulse. Both approaches converge to a $G$ error = 0.014 and $G'$ error = 0.28 ($G'$ is normalized by the trace area, rather than the number of points and so can be compared to the trace noise). Both $G$ values indicate excellent agreement, as does visual inspection of the traces. (e,f) retrieved temporal and spectral intensity and phase, respectively.

## 6. Discussion

The RANA approach involves directly retrieving the pulse spectrum from the SHG FROG trace, defining multiple initial guesses with that retrieved spectrum and random spectral phases, and using multiple grids. In our study, this approach converged for every single one of the more than 25,000 pulses we tried, including those with extremely large *TBP*s of 100. For extremely complex pulses, the RANA approach also converged much faster than the standard generalized projections algorithm—which we continue to use in the RANA approach, but within the framework of the multiple initial guesses, essentially all having the correct spectra, and the multiple grids.

The high reliability of the RANA approach is due mainly to the use of vastly improved initial guesses involving the correct (or nearly correct) spectrum (and spectral phase). This was evident because, if the retrieved spectrum from the frequency marginal was used as the initial guess, the standard GP algorithm's convergence performance for only one initial guess on the full grids increases to well above 90% (especially for the large *TBP*s, interestingly). See Table 4, which compares the performance of the GP algorithm with a random complex initial guess and an initial guess with spectral amplitude obtained from the frequency marginal and random spectral phase. The effect of the improved initial guess can be seen by the significant improvement in the percentage of convergence of GP algorithm.

Table 4. Performance of standard GP algorithm with an initial guess of the spectral intensity obtained from the frequency marginal of SHG FROG trace.

| Initial guess | Pulse *TBP* | Percentage of convergence | Average retrieval time |
|---|---|---|---|
| Random complex array | 2.5 | 75% | 0.051s |
| Retrieved spectrum and random phase | | 92% | 0.041s |
| Random complex array | 5 | 74% | 0.30s |
| Retrieved spectrum and random phase | | 95% | 0.20s |
| Random complex array | 10 | 70% | 1.29s |
| Retrieved spectrum and random phase | | 98% | 0.85s |
| Random complex array | 20 | 63% | 6.03s |
| Retrieved spectrum and random phase | | 98% | 4.37s |
| Random complex array as initial guess | 40 | 56% | 45.7s |
| Retrieved spectrum and random phase | | 97% | 27.1s |

The multiple initial guesses increase reliability as well but only occasionally prove necessary.

The multiple (that is, smaller) grids serve mainly to speed convergence.

The use of parallel processing is so trivially implemented that multiple initial guesses would seem obviously useful to all FROG algorithms. Parallelism also speeds convergence, of course. For example, for pulses with *TBP* = 5, in the absence of the implementation of parallelism, the average retrieval time approximately doubled. Of course, MATLAB's fast Fourier transform also uses parallel processing, so the standard GP algorithm with only one guess benefits from it as well, but some steps in it do not naturally lend themselves to parallel processing. The RANA approach, on the other hand, does so more completely and at every step and so operated more rapidly for complex pulses.

As the four or eight remaining guesses were considered consecutively for the final few iterations in the final grid (the full $N \times N$ trace), and convergence was nearly always achieved for the first initial guess tried, it is likely that fewer initial guesses are in fact necessary. For example, because the mere use of the correct (or nearly correct) spectrum for the initial guess yields a >90% probability of convergence in the standard GP approach, use of twenty initial guesses would yield a probability of stagnation of $10^{-20}$ if independent trials can be assumed. This implies that the RANA approach is vastly more powerful than is actually necessary for the problem and hence includes much flexibility in the trade-off between reliability and speed. We have not optimized all the approach's parameters, and it would seem that increased speed beyond our results is likely without sacrificing robustness.

It is interesting that the RANA approach works so well even in the presence of noise, when the spectrum-retrieval process occasionally failed. This is likely because the retrieved spectrum, even in these cases, is much closer to the correct spectrum than currently used initial guesses consisting entirely of random noise or random temporal phase noise in conjunction with the autocorrelation approximating the temporal intensity. It should also be mentioned that we have not fully taken advantage of the positivity of spectrum, as the occasional spectral negativity can be used for the elimination of the incorrect sign of $s_{\pm}(t)$ when the values of $\varepsilon_{\pm}$ are close to zero and close to each other. In our study, we have only used the symmetry properties of the real and imaginary parts and obtained the spectrum from two halves of data (by considering evenness and oddness of real and imaginary parts, respectively) and rejected the one that yields the larger number of negative points. Furthermore, as mentioned in the text, there are alternative approaches that could be implemented to directly retrieve the spectrum rather than the weighted sums; for example, a sequential comparison of first continuity, then the first derivative, then the second derivative, etc.

Finally, we used MATLAB for this comparative study, but much faster programming languages are available, and convergence should be considerably faster using such a language.

## 7. Conclusion

In this work, we implement an algorithmic approach for the SHG FROG pulse-retrieval algorithm that benefits from significantly better initial guesses that are obtained directly from the SHG trace. These initial guesses contain the spectral intensity of the pulse that is obtained from the frequency marginal of SHG trace based on Paley-Wiener Theorem. Furthermore, by implementation of a multi-grid scheme and multiple initial guesses, we achieved 100% convergence, which was also faster for more complicated pulses. Overall, the RANA approach is quite general: if an improved (say, at least as reliable, but faster) single-guess, single-grid algorithm emerges someday, it could also be used within the RANA approach in place of the current standard generalized projections algorithm, thus achieving even faster convergence. With the RANA approach, ultrahigh reliability in SHG FROG pulse retrieval has finally arrived.

**Funding**

National Science Foundation (#ECCS-1307817) and the Georgia Research Alliance.

**Acknowledgments**

We would like to thank Esmerando Escoto for the helpful discussions.

**Disclosures**

Rick Trebino owns a company that sells pulse-measurement devices.